\documentclass{article}

\PassOptionsToPackage{numbers,sort&compress}{natbib} % numeric citations, NeurIPS standard
\usepackage[preprint]{neurips_2026} 

\usepackage[utf8]{inputenc} % allow utf-8 input
\usepackage[T1]{fontenc}    % use 8-bit T1 fonts
\usepackage[colorlinks=true, citecolor=blue, linkcolor=blue, urlcolor=blue]{hyperref} % hyperlinks
\usepackage{url}            % simple URL typesetting
\usepackage{booktabs}       % professional-quality tables
\usepackage{amsfonts}       % blackboard math symbols
\usepackage{nicefrac}       % compact symbols for 1/2, etc.
\usepackage{microtype}      % microtypography
\usepackage{xcolor}         % colors
\usepackage{graphicx}
\usepackage{placeins}
\usepackage{booktabs,tabularx,multirow,array}
\usepackage{float}          % [H] figure placement
\usepackage{tikz}
\usepackage{threeparttable}   % in preamble
\usepackage{amsmath}
\usepackage{subcaption}
\usepackage[numbers,sort&compress]{natbib}
\usetikzlibrary{arrows.meta, decorations.pathreplacing, calc}

\definecolor{loc}{HTML}{C2410C}
\definecolor{bio}{HTML}{0E7490}
\definecolor{cyber}{HTML}{6D28D9}
\definecolor{pc}{HTML}{A16207}
\definecolor{ink}{HTML}{1A1A1A}
\definecolor{mid}{HTML}{444444}
\definecolor{faint}{HTML}{999999}
\definecolor{coregray}{HTML}{4B5563}
\definecolor{insgray}{HTML}{9CA3AF}

\title{Capability-Based Planning \\for AI Crisis Preparedness}
\author{%
  Isaak Mengesha$^{1,2}$\thanks{Corresponding author} \And
  Charlie Collins$^{1}$ \And
  Juan Felipe Cerón Uribe$^{1}$ \And
  Salvatore d'Ambrosio$^{1}$ \And
  Vickie Ellis$^{1}$ 
  \AND
  \normalfont
  $^{1}$Arcadia Impact AI Governance Taskforce \quad
  $^{2}$University of Oxford \quad
}

\begin{document}

\maketitle

\begin{abstract}
Capability-based planning drives preparedness in defense and homeland security, but has yet to be applied seriously to AI. Government AI preparations follow a predict-then-act paradigm: rank risks by likelihood and impact, then prepare for the highest expected harm. AI resists prediction: expert timelines disagree by orders of magnitude, and official reviews concede that likelihood-based risk assessment fails for exactly this class of risk. Drawing on principles of decision making under deep uncertainty, we propose a methodological framework in three parts: a scenario library sampled systematically across declared axes; a rating procedure that assesses each government capability against each scenario on coarse, gated criteria; and a prioritization step that maps the resulting matrix onto decision rules a government might adopt. Through a pilot across the four most severe AI-enabled threat classes, we illustrate the kind of insight the instrument yields and provide a proof of concept for capability-based planning as a practical tool for AI crisis preparedness.
\end{abstract}

\section{Introduction}\label{sec:intro}

%\section*{§1: The range of possible AI futures will outpace and outflank preparations based on a narrow ``predict-then-act'' paradigm}

The future of advanced artificial intelligence (AI) is highly uncertain, spanning a broad range of possible paths, some of which pose severe risks. Across jurisdictions, government preparations include: forming and connecting AI safety institutes \cite{barnett2025ai, allen2024aisi}; encouraging (and sometimes requiring) frontier safety frameworks \citep{fmf2025frameworks, ca2025sb53, ec2025gpaicop}; mandating serious-incident reporting \citep{euaiact2024, ny2026raiseact}; and updating national risk registers \citep{cabinetoffice2025nrr} and incident-response plans \citep{whitehouse2025actionplan}. These preparations generally follow a "predict-then-act" paradigm \cite{marchau2019dmdu}: risk registers rank threats by likelihood and impact \cite{cabinetoffice2025nrr,blagden2018nsra}, and AI safety institutes conduct pre-deployment tests for dangerous capabilities \cite{barnett2025ai, allen2024aisi}. Prevention and mitigation policies are based on a relatively small number of specific threat models, with the implicit assumption that emergent AI threats can be identified, accurately ranked and met with a suitable strategy before they cause harm \citep{mengesha2026coordination, boudreaux2025loc, wasil2024emergency}. However, AI resists prediction: experts disagree by orders of magnitude on AI timelines \citep{grace2024thousands}, capabilities can emerge abruptly \cite{wei2022emergent}, and failure modes are unpredictable \cite{ganguli2022predictability}.

%need more of a link between these paragraphs

Government capabilities designed for a specific scenario rarely generalise well \citep{davis2002cbp, rusi2020nrr}, and even the most probable scenario is individually unlikely to arrive as predicted---or at all---making narrow, scenario-specific investments a poor bet \citep{lempert2003rdm}. For example, an influenza pandemic topped the UK's risk register from 2008-2020, but preparations failed to anticipate the asymptomatic transmission that defined COVID-19, rendering the response ineffective \citep{covidinquiry2024m1,rusi2020nrr}. An official review recommended preparing for a wider range of scenarios and paying closer attention to interdependencies \citep{raeng2023resilience}.

% FLAGGED FOR CUTTING (CONDENSED VERSION INCLUDED ABOVE): A post-COVID-19 review of National Security Risk Assessment (NSRA) methodology recommended the introduction of multiple-scenario design, attention to interdependencies, and investment in rapidly deployable critical capabilities \citep{raeng2023resilience}.

Thresholds for action also entail problems: disagreement over where they lie, or whether they have been crossed, can block coordinated action. This is demonstrated in climate negotiations \cite{barrett2014sensitivity} and is equally relevant to AI capability thresholds meant to trigger safety measures \cite{morris2024levels}. The UK's National Security Risk Assessment (NSRA) excludes emerging risks beyond a two-year horizon, or below an assessed 1-in-100,000-year likelihood; these are the kind of high-uncertainty, novel risks posed by advanced AI \citep{hlords2021extreme,blagden2018nsra}, which some forecasts now place well above that likelihood \citep{karger2023xpt, delphi2026risks}.

%CHARLIE NOTE: THIS PARAGRAPH REPLACED BY MATERIAL BELOW: Alternatives to forecast-led planning already exist. "If-then commitments" (actions pre-agreed against observed conditions rather than against predictions of when those conditions will arrive) have been adopted, if unevenly, across frontier-labs  \citep{metr2025common, koessler2024thresholds}. But no equivalent governs how the state itself would respond, and that absence is what allows voluntary restraint to unravel: Anthropic's revised commitment now holds only if competitors verifiably match it \citep{anthropic2026rsp, time2026anthropic}, so no developer restrains itself alone when doing so cedes ground. A binding state response applied to all developers at once --- a mandated model withdrawal once a defined risk materialises, say --- changes the calculus, turning restraint from a competitive sacrifice into a shared baseline. To achieve this, states need to focus on capabilities and actions instead of prediction.

%\section*{§2: Voluntary ``if-then'' commitments are a partial fix — the state needs the same capability, and the power to enforce it}

``If-then commitments,'' adopted voluntarily by frontier AI developers, partially escape this reliance on prediction. Predefined thresholds for dangerous capabilities (the "if") trigger specific safety, security, or governance measures (the "then"), fixing who acts and how, without needing to predict when \citep{metr2025common, koessler2024thresholds}. Under this paradigm, evaluation regimes act as tripwires rather than forecasts.

First proposed with nationstates in mind \citep{karnofsky2024ifthen}, the same principle applied at national scale demands capabilities commensurate with the state's remit. States need to know---well before a threshold is crossed---who will act, and with what authority, resources, and expertise.

States should also be empowered to harmonise, verify, and enforce developers' own if-then commitments. Currently, published thresholds differ substantially between developers, with no reliable means to confirm whether one has been crossed \citep{anterola2026harmonizing}. Without verification and enforcement, commercial pressure to keep pace with rivals tends to trump voluntary restraint \citep{anthropic2026rsp, time2026anthropic}.

%\section*{§3: Governments already ``monitor-and-adapt'' to build resilience under uncertainty---they should apply this experience to AI}

% %CHARLIE NOTE: THESE TWO PARAGRAPHS REPLAED BY MATERIAL BELOW: Fields that moved past point-forecasts did not give up on planning. Instead of asking \textit{which} future, they asked what must hold \textit{across} futures \citep{lempert2003rdm}. Post-cold-war defense, flood policy, and post-2008 financial supervision all made the same move: from preparing against a predicted threat to building capabilities robustly across scenarios \citep{davis2002cbp, haasnoot2013dapp, frb2009scap}.
% The UK Government Resilience Action Plan now follows this model, supplementing likelihood-and-impact ranking with an assurance layer that defines what the state must be able to do: a National Capabilities Assessment scrutinising response capabilities across government, strengthened through independent red-teaming \citep{hmg2025rap}. Applying the same capability lens to AI would not ask government to invent a posture, only to extend one it is already adopting for conventional risk into the domain where likelihood estimates are least reliable, and where, as external reviewers note, acute AI risks remain largely absent from the national register \citep{cltr2025rap}.

Testing what capabilities must hold across many possible futures, rather than betting on which future is likeliest, is not new to policymaking, but has yet to be applied seriously to AI. Developed in the early-to-mid 2000s and formalised in the field of Decision Making under Deep Uncertainty (DMDU) \citep{marchau2019dmdu}, the "monitor-and-adapt" paradigm shapes such diverse fields as US defence and financial supervisory planning, and Dutch flood mitigation \citep{lempert2003rdm, davis2002cbp, haasnoot2013dapp, frb2009scap}. In the UK, the Resilience Action Plan follows a similar paradigm by supplementing the NSRA's likelihood-and-impact ranking with a capability-assurance layer: a National Capabilities Assessment (NCA) that tests government-wide response capability \citep{hmg2025rap}.

 %FLAGGED FOR CUTTING AS EXTRANEOUS/REPETITIVE: although the UK's National Risk Register largely excludes acute AI risks \citep{cltr2025rap}.

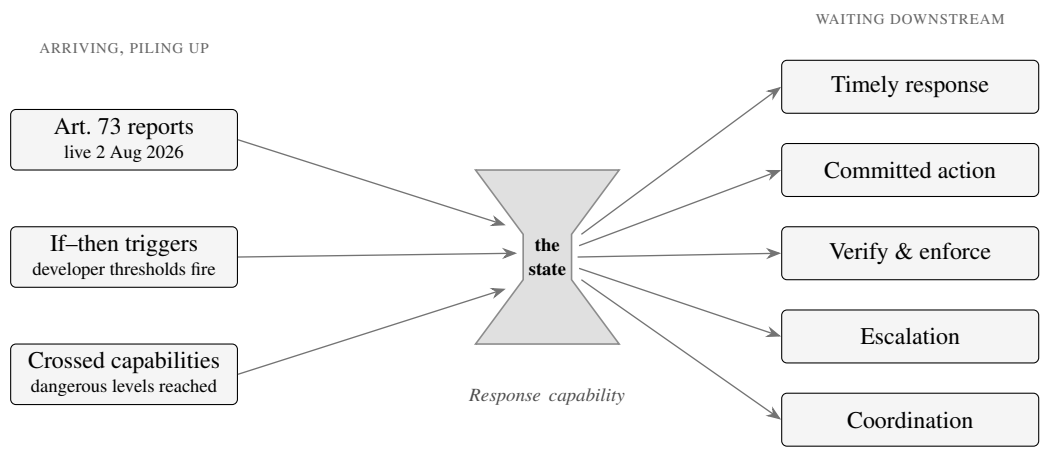
\begin{figure}[H]
  \centering
  % ---------------------------------------------------------------
% The unprepared state as a bottleneck.
% Requires in the preamble of the master document:
%   \usepackage{tikz}
%   \usetikzlibrary{arrows.meta,positioning,calc,shapes.geometric}
% Colours are defined locally below so this file is self-contained.
% ---------------------------------------------------------------
\begin{tikzpicture}[
    font=\small,
    >={Stealth[length=2mm]},
    inflow/.style   = {draw, rounded corners=2pt, fill=black!4,
                       minimum width=30mm, minimum height=8mm,
                       align=center, inner sep=2pt, line width=0.4pt},
    demand/.style   = {draw, rounded corners=2pt, fill=black!4,
                       minimum width=34mm, minimum height=7mm,
                       align=center, inner sep=2pt, line width=0.4pt},
    flow/.style     = {->, line width=0.5pt, black!55},
  ]

  % ---- inflow (left): pressure arriving and piling up ----
  \node[inflow] (rep)  at (0,1.55)  {Art.\ 73 reports\\[-1pt]{\scriptsize live 2 Aug 2026}};
  \node[inflow] (trig) at (0,0)     {If--then triggers\\[-1pt]{\scriptsize developer thresholds fire}};
  \node[inflow] (cap)  at (0,-1.55) {Crossed capabilities\\[-1pt]{\scriptsize dangerous levels reached}};

  % ---- the choke (centre): a physically narrow gate ----
  % drawn as an hourglass-style pinch so the constriction is visible
  \def\cx{5.6}
  \fill[black!12, draw=black!45, line width=0.6pt]
      (\cx-0.95,1.15) -- (\cx-0.32,0.30) -- (\cx-0.32,-0.30) -- (\cx-0.95,-1.15) --
      (\cx+0.95,-1.15) -- (\cx+0.32,-0.30) -- (\cx+0.32,0.30) -- (\cx+0.95,1.15) -- cycle;
  \node[align=center] at (\cx,0) {\scriptsize\bfseries the\\[-1pt]\scriptsize\bfseries state};
  \node[align=center, text width=42mm, black!70] at (\cx,-1.85)
      {\scriptsize\emph{Response capability}\\[-2pt]{\scriptsize }};

  % ---- demand (right): everything waiting downstream ----
  \node[demand] (r1) at (10.4,2.25)  {Timely response};
  \node[demand] (r2) at (10.4,1.15)  {Committed action};
  \node[demand] (r3) at (10.4,0.05)  {Verify \& enforce};
  \node[demand] (r4) at (10.4,-1.05) {Escalation};
  \node[demand] (r5) at (10.4,-2.15) {Coordination};

  % ---- flows in: converging to the pinch ----
  \draw[flow] (rep.east)  -- (\cx-0.55,0.42);
  \draw[flow] (trig.east) -- (\cx-0.40,0.05);
  \draw[flow] (cap.east)  -- (\cx-0.55,-0.42);

  % ---- flows out: fanning from the pinch, thin (throttled) ----
  \draw[flow] (\cx+0.45,0.30)  -- (r1.west);
  \draw[flow] (\cx+0.42,0.15)  -- (r2.west);
  \draw[flow] (\cx+0.40,0.00)  -- (r3.west);
  \draw[flow] (\cx+0.42,-0.15) -- (r4.west);
  \draw[flow] (\cx+0.45,-0.30) -- (r5.west);

  % ---- column headers ----
  \node[black!60] at (0,2.75)    {\scriptsize\textsc{arriving, piling up}};
  \node[black!60] at (10.4,3.15) {\scriptsize\textsc{waiting downstream}};

\end{tikzpicture}
  \caption{\textbf{State capacity as the bottleneck in crises and pivotal moments.} Private and, increasingly, public governance systems are coming into operation. They will not prevent crises entirely, but they increasingly demarcate points of action and response needs. This places more, not less, demand on nation states to respond appropriately.}
  \label{fig:bottleneck}
\end{figure}

% \section*{§4: Government can specify and prioritie future capabilities based on existing risk research}

%\section*{§4: Governments can use current AI risk research to specify and prioritise capabilities robust to an uncertain AI future}

% %CHARLIE NOTE: THis PARAGRAPH IS REPLACED BY MATERIAL BELOW:  While risk probability is an exercise in uncertainty, a government's required response capabilities are more easily defined. A scenario in which a frontier model self-exfiltrates and keeps operating across rented cloud compute already determines who must be able to detect it, which agency owns the response, who is consulted, and what legal authorities --- such as compelled access to compute providers, mandated shutdown --- the response requires. This is the shift from the capability side of AI to the capabilities of government for which the adaptation literature argues \citep{bernardi2024adaptation}. No existing resource maps risk to government capabilities. Incident databases and risk taxonomies sort failures by cause and harm, not by necessary action \citep{slattery2024repository, aiid}. The preparedness work that exists --- policy lists with emergency powers \citep{cltr2025incidents}, a loss-of-control response plan \citep{boudreaux2025loc}, federal gap assessments \citep{wasil2024emergency}, international crisis protocols \citep{chathamhouse2026deadlock} --- covers single threat classes or hand-picked scenarios, derives no \textit{robust} capability set from a stated scenario space, and consequently struggles to prioritize.\\

Defining government response capability is tractable across a wide range of uncertain futures, while accurately assessing and comparing risks is not \citep{bernardi2024adaptation}. Envisioning a possible future scenario in which, for example, an unreleased frontier model self-exfiltrates to rented cloud compute is enough to determine---if it happened today---who would detect it, which agency would own it, who should be consulted, and what legal powers would be required to stop it (e.g. access to compute providers and mandated shutdown). However, no existing resource systematically maps a wide range of plausible AI scenarios to the required response capability. Incident databases and risk taxonomies sort failures by cause and harm, not remedy \citep{slattery2024repository, aiid}, and preparedness work targets single threats or hand-picked scenarios: emergency powers, loss-of-control response plans, federal capability-gap assessments, international crisis protocols \citep{cltr2025incidents, boudreaux2025loc, wasil2024emergency, chathamhouse2026deadlock}.

% CHARLIE NOTE: THis PARAGRAPH IS REPLACED BY MATERIAL BELOW: FEMA's Target Capabilities List and the UK's National Capabilities Assessment show that derivation is feasible \citep{dhs2007tcl}(§3). Deep uncertainty changes only the input: a broad, openly declared set of scenarios that anyone can challenge and extend helps ensure that the capability list remains agnostic to any single author's guess about the future \citep{lempert2003rdm}. The capabilities demanded across any such set exceed any budget, so prioritization happens with or without probabilities. Refusing explicit prioritization delegates it to inertia, salience and bias (the implicit mode the pandemic record of §2 exemplifies: ranking without action-centered analysis) --- what is important is whether the process is accountable. Prioritization of government capabilities could occur along two strategies:

The US Federal Emergency Management Agency's (FEMA) Target Capabilities List and the UK's NCA show that governments can already perform the significant task of mapping scenarios to capabilities at national scale \citep{dhs2007tcl, hmg2025rap}. Applying this to AI demands little new: governments need only develop a sufficiently broad, representative set of AI scenarios. Consideration of such a set will produce an extensive list of capabilities well beyond realistic resource constraints. Two criteria can guide robust capability prioritisation under deep uncertainty: value across many possible futures, and low-cost protection against the most severe ones \citep{marchau2019dmdu, haasnoot2013dapp}. Both steps---generating the scenario set and choosing among the resulting capabilities---should follow an explicit, accountable, and recurring process, revisited whenever monitored conditions cross pre-agreed signposts \citep{haasnoot2013dapp}. Robust Decision Making (RDM) provides one such iterative, stakeholder-inclusive method \citep{lempert2003rdm}. Without it, government risks defaulting to bureaucratic inertia: reacting to whichever risk is currently salient \citep{baumgartner1993agendas}, or to a narrow or biased range of possible futures.

\section{Methods}
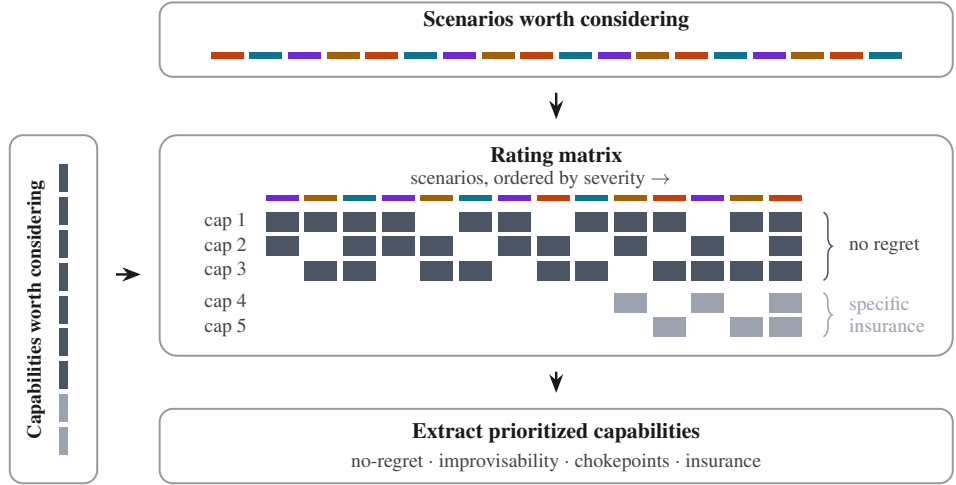
\begin{figure}[H]
  \centering
    \resizebox{0.9\textwidth}{!}{\begin{tikzpicture}[
  yscale=0.85,
  axislabel/.style={mid, font=\normalsize},
  famlabel/.style={font=\normalsize\bfseries},
  paneltitle/.style={ink, font=\large\bfseries, anchor=west},
  gridtext/.style={mid, font=\small},
]

%% ================= SCENARIOS INPUT (chips) =================
\draw[faint, line width=0.9pt, rounded corners=6pt] (1.2,5.85) rectangle (13.9,7.25);
\node[ink, font=\normalsize\bfseries] at (7.55,6.90) {Scenarios worth considering};
\fill[loc] (2.02,6.19) rectangle (2.54,6.32);
\fill[bio] (2.64,6.19) rectangle (3.16,6.32);
\fill[cyber] (3.26,6.19) rectangle (3.78,6.32);
\fill[pc] (3.88,6.19) rectangle (4.40,6.32);
\fill[loc] (4.50,6.19) rectangle (5.02,6.32);
\fill[bio] (5.12,6.19) rectangle (5.64,6.32);
\fill[cyber] (5.74,6.19) rectangle (6.26,6.32);
\fill[pc] (6.36,6.19) rectangle (6.88,6.32);
\fill[loc] (6.98,6.19) rectangle (7.50,6.32);
\fill[bio] (7.60,6.19) rectangle (8.12,6.32);
\fill[cyber] (8.22,6.19) rectangle (8.74,6.32);
\fill[pc] (8.84,6.19) rectangle (9.36,6.32);
\fill[loc] (9.46,6.19) rectangle (9.98,6.32);
\fill[bio] (10.08,6.19) rectangle (10.60,6.32);
\fill[cyber] (10.70,6.19) rectangle (11.22,6.32);
\fill[pc] (11.32,6.19) rectangle (11.84,6.32);
\fill[loc] (11.94,6.19) rectangle (12.46,6.32);
\fill[bio] (12.56,6.19) rectangle (13.08,6.32);

\draw[ink, line width=1.1pt, -{Stealth[length=3mm]}] (7.55,5.55) -- (7.55,5.05);

%% ================= CAPABILITIES INPUT =================
\draw[faint, line width=0.9pt, rounded corners=6pt] (-1.2,-1.80) rectangle (0.2,4.75);
\node[ink, font=\small\bfseries, rotate=90] at (-0.78,1.475) {Capabilities worth considering};
\fill[insgray] (-0.40,-1.27) rectangle (-0.27,-0.75);
\fill[insgray] (-0.40,-0.65) rectangle (-0.27,-0.13);
\fill[coregray] (-0.40,-0.03) rectangle (-0.27,0.49);
\fill[coregray] (-0.40,0.59) rectangle (-0.27,1.11);
\fill[coregray] (-0.40,1.21) rectangle (-0.27,1.73);
\fill[coregray] (-0.40,1.83) rectangle (-0.27,2.35);
\fill[coregray] (-0.40,2.45) rectangle (-0.27,2.97);
\fill[coregray] (-0.40,3.07) rectangle (-0.27,3.59);
\fill[coregray] (-0.40,3.69) rectangle (-0.27,4.21);
\draw[ink, line width=1.1pt, -{Stealth[length=3mm]}] (0.50,2.13) -- (0.90,2.13);

%% ================= RATING MATRIX =================
\draw[faint, line width=0.9pt, rounded corners=8pt] (1.2,0.60) rectangle (13.9,4.75);
\node[ink, font=\normalsize\bfseries] at (7.55,4.35) {Rating matrix};
\node[gridtext, anchor=south] at (7.30,3.60) {scenarios, ordered by severity $\rightarrow$};
\fill[cyber] (2.90,3.52) rectangle (3.42,3.62);
\fill[pc] (3.52,3.52) rectangle (4.04,3.62);
\fill[bio] (4.14,3.52) rectangle (4.66,3.62);
\fill[cyber] (4.76,3.52) rectangle (5.28,3.62);
\fill[pc] (5.38,3.52) rectangle (5.90,3.62);
\fill[bio] (6.00,3.52) rectangle (6.52,3.62);
\fill[cyber] (6.62,3.52) rectangle (7.14,3.62);
\fill[loc] (7.24,3.52) rectangle (7.76,3.62);
\fill[bio] (7.86,3.52) rectangle (8.38,3.62);
\fill[pc] (8.48,3.52) rectangle (9.00,3.62);
\fill[loc] (9.10,3.52) rectangle (9.62,3.62);
\fill[cyber] (9.72,3.52) rectangle (10.24,3.62);
\fill[pc] (10.34,3.52) rectangle (10.86,3.62);
\fill[loc] (10.96,3.52) rectangle (11.48,3.62);
\node[anchor=east, gridtext] at (2.72,3.12) {cap 1};
\node[anchor=east, gridtext] at (2.72,2.66) {cap 2};
\node[anchor=east, gridtext] at (2.72,2.20) {cap 3};
\node[anchor=east, gridtext] at (2.72,1.60) {cap 4};
\node[anchor=east, gridtext] at (2.72,1.14) {cap 5};
\fill[coregray] (2.90,2.94) rectangle (3.42,3.30);
\fill[coregray] (3.52,2.94) rectangle (4.04,3.30);
\fill[coregray] (4.14,2.94) rectangle (4.66,3.30);
\fill[coregray] (4.76,2.94) rectangle (5.28,3.30);
\fill[coregray] (6.00,2.94) rectangle (6.52,3.30);
\fill[coregray] (6.62,2.94) rectangle (7.14,3.30);
\fill[coregray] (7.86,2.94) rectangle (8.38,3.30);
\fill[coregray] (8.48,2.94) rectangle (9.00,3.30);
\fill[coregray] (9.10,2.94) rectangle (9.62,3.30);
\fill[coregray] (10.34,2.94) rectangle (10.86,3.30);
\fill[coregray] (10.96,2.94) rectangle (11.48,3.30);
\fill[coregray] (2.90,2.48) rectangle (3.42,2.84);
\fill[coregray] (4.14,2.48) rectangle (4.66,2.84);
\fill[coregray] (4.76,2.48) rectangle (5.28,2.84);
\fill[coregray] (5.38,2.48) rectangle (5.90,2.84);
\fill[coregray] (6.62,2.48) rectangle (7.14,2.84);
\fill[coregray] (7.24,2.48) rectangle (7.76,2.84);
\fill[coregray] (8.48,2.48) rectangle (9.00,2.84);
\fill[coregray] (9.72,2.48) rectangle (10.24,2.84);
\fill[coregray] (10.96,2.48) rectangle (11.48,2.84);
\fill[coregray] (3.52,2.02) rectangle (4.04,2.38);
\fill[coregray] (4.14,2.02) rectangle (4.66,2.38);
\fill[coregray] (5.38,2.02) rectangle (5.90,2.38);
\fill[coregray] (6.00,2.02) rectangle (6.52,2.38);
\fill[coregray] (7.24,2.02) rectangle (7.76,2.38);
\fill[coregray] (7.86,2.02) rectangle (8.38,2.38);
\fill[coregray] (9.10,2.02) rectangle (9.62,2.38);
\fill[coregray] (9.72,2.02) rectangle (10.24,2.38);
\fill[coregray] (10.34,2.02) rectangle (10.86,2.38);
\fill[coregray] (10.96,2.02) rectangle (11.48,2.38);
\fill[insgray] (8.48,1.42) rectangle (9.00,1.78);
\fill[insgray] (9.72,1.42) rectangle (10.24,1.78);
\fill[insgray] (10.96,1.42) rectangle (11.48,1.78);
\fill[insgray] (9.10,0.96) rectangle (9.62,1.32);
\fill[insgray] (10.34,0.96) rectangle (10.86,1.32);
\fill[insgray] (10.96,0.96) rectangle (11.48,1.32);
\draw[coregray, line width=0.7pt, decorate, decoration={brace, amplitude=4pt}]
  (11.83,3.30) -- node[gridtext, anchor=west, xshift=8pt] {no regret} (11.83,2.02);
\draw[insgray, line width=0.7pt, decorate, decoration={brace, amplitude=4pt}]
  (11.83,1.78) -- node[insgray, font=\small, anchor=west, xshift=8pt, align=left] {specific\\ insurance} (11.83,0.96);

%% ================= EXTRACTION =================
\draw[ink, line width=1.1pt, -{Stealth[length=3mm]}] (7.55,0.30) -- (7.55,-0.10);
\draw[faint, line width=0.9pt, rounded corners=6pt] (1.2,-1.80) rectangle (13.9,-0.40);
\node[ink, font=\normalsize\bfseries] at (7.55,-0.85) {Extract prioritized capabilities};
\node[mid, font=\small] at (7.55,-1.40) {no-regret $\cdot$ improvisability $\cdot$ chokepoints $\cdot$ insurance};

\end{tikzpicture}}\caption{\textbf{A high-level outline of the approach.} Scenarios and capabilities are crossed into a rating matrix; applying a decision rule extracts a prioritized set of capabilities.}\label{fig:overview}
\end{figure}

Here we apply the principles of RDM to the landscape of extreme risk from transformative AI. The broader overview, shown in figure \ref{fig:overview}, decomposes into the principles that determine the inclusion of scenarios and capabilities into an assessment matrix, which in turn allows us to extract prioritisation with regard to defined strategies. 

\paragraph{Selecting Scenarios.} In our subsequent work, we focus on four broader scenario classes \footnote{power concentration, loss of control, AI\,$\times$\,bio, AI\,$\times$\,cyber \citep{koessler2024thresholds, metr2025common, boudreaux2025loc, wasil2024emergency}} discussed in the literature as particularly concerning \citep{bengio2026international, delphi2026risks, slattery2024repository}. Where possible, we opted for classes/scenarios with rather clear trigger events or tangible harms as they more easily define crisis onset and points of intervention \citep{kulveit2025gradual}. Within each class we collect scenarios with the intention to maximize variance of scenarios represented, rather than likelihood (e.g. expected harm) to acknowledge our deep uncertainty \citep{davis2002cbp, lempert2003rdm, groves2007new, bankes1993exploratory}.\\

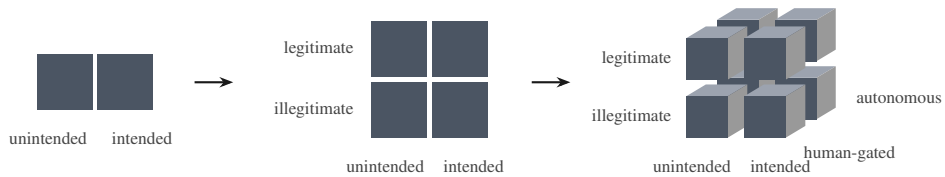
\begin{figure}[H]
  \centering
    \resizebox{0.9\textwidth}{!}{\begin{tikzpicture}[
  axlab/.style={mid, font=\small},
  axname/.style={ink, font=\small\itshape},
  flow/.style={ink, line width=1.1pt, -{Stealth[length=2.5mm]}},
]

%% ---------- Panel 1 ----------
\fill[coregray] (0.70,1.40) rectangle (1.70,2.40);
\fill[coregray] (1.80,1.40) rectangle (2.80,2.40);
\node[axlab, anchor=north] at (0.90,1.12) {unintended};
\node[axlab, anchor=north] at (2.60,1.12) {intended};

\draw[flow] (3.55,1.90) -- (4.25,1.90);

%% ---------- Panel 2 ----------
\fill[coregray] (6.75,0.90) rectangle (7.75,1.90);
\fill[coregray] (6.75,2.00) rectangle (7.75,3.00);
\fill[coregray] (7.85,0.90) rectangle (8.85,1.90);
\fill[coregray] (7.85,2.00) rectangle (8.85,3.00);
\node[axlab, anchor=north] at (7.00,0.62) {unintended};
\node[axlab, anchor=north] at (8.60,0.62) {intended};
\node[axlab, anchor=east] at (6.55,1.40) {illegitimate};
\node[axlab, anchor=east] at (6.55,2.50) {legitimate};

\draw[flow] (9.65,1.90) -- (10.35,1.90);

%% ---------- Panel 3 ----------
\fill[insgray] (13.00,3.03) -- (13.35,3.24) -- (14.10,3.24) -- (13.75,3.03) -- cycle;
\fill[faint] (13.75,2.28) -- (14.10,2.49) -- (14.10,3.24) -- (13.75,3.03) -- cycle;
\fill[coregray] (13.00,2.28) rectangle (13.75,3.03);
\fill[insgray] (14.05,3.03) -- (14.40,3.24) -- (15.15,3.24) -- (14.80,3.03) -- cycle;
\fill[faint] (14.80,2.28) -- (15.15,2.49) -- (15.15,3.24) -- (14.80,3.03) -- cycle;
\fill[coregray] (14.05,2.28) rectangle (14.80,3.03);
\fill[insgray] (13.00,1.98) -- (13.35,2.19) -- (14.10,2.19) -- (13.75,1.98) -- cycle;
\fill[faint] (13.75,1.23) -- (14.10,1.44) -- (14.10,2.19) -- (13.75,1.98) -- cycle;
\fill[coregray] (13.00,1.23) rectangle (13.75,1.98);
\fill[insgray] (14.05,1.98) -- (14.40,2.19) -- (15.15,2.19) -- (14.80,1.98) -- cycle;
\fill[faint] (14.80,1.23) -- (15.15,1.44) -- (15.15,2.19) -- (14.80,1.98) -- cycle;
\fill[coregray] (14.05,1.23) rectangle (14.80,1.98);
\fill[insgray] (12.45,2.70) -- (12.80,2.91) -- (13.55,2.91) -- (13.20,2.70) -- cycle;
\fill[faint] (13.20,1.95) -- (13.55,2.16) -- (13.55,2.91) -- (13.20,2.70) -- cycle;
\fill[coregray] (12.45,1.95) rectangle (13.20,2.70);
\fill[insgray] (13.50,2.70) -- (13.85,2.91) -- (14.60,2.91) -- (14.25,2.70) -- cycle;
\fill[faint] (14.25,1.95) -- (14.60,2.16) -- (14.60,2.91) -- (14.25,2.70) -- cycle;
\fill[coregray] (13.50,1.95) rectangle (14.25,2.70);
\fill[insgray] (12.45,1.65) -- (12.80,1.86) -- (13.55,1.86) -- (13.20,1.65) -- cycle;
\fill[faint] (13.20,0.90) -- (13.55,1.11) -- (13.55,1.86) -- (13.20,1.65) -- cycle;
\fill[coregray] (12.45,0.90) rectangle (13.20,1.65);
\fill[insgray] (13.50,1.65) -- (13.85,1.86) -- (14.60,1.86) -- (14.25,1.65) -- cycle;
\fill[faint] (14.25,0.90) -- (14.60,1.11) -- (14.60,1.86) -- (14.25,1.65) -- cycle;
\fill[coregray] (13.50,0.90) rectangle (14.25,1.65);
\node[axlab, anchor=north] at (12.55,0.62) {unintended};
\node[axlab, anchor=north] at (14.15,0.62) {intended};
\node[axlab, anchor=east] at (12.30,1.28) {illegitimate};
\node[axlab, anchor=east] at (12.30,2.33) {legitimate};
\node[axlab, anchor=north] at (15.40,0.87) {human-gated};
\node[axlab, anchor=west] at (15.40,1.62) {autonomous};

\end{tikzpicture}}
  \caption{\textbf{Transparent variance maximization.} Scenarios are sampled at every combination of three precommitted binary axes, with undeclared dimensions as potentially biased. Axis definitions are given in Table~\ref{tab:dim}.}\label{fig:cubeviz}
\end{figure}

While there is no clear limit to the number of scenarios to include in principle, introducing them without curation risks introducing bias. For example, 12 of the 15 FEMA National Planning Scenarios were terrorism-related, skewing the resulting capability list toward those threats at the expense of others \citep{dhs2007tcl}. It may be impossible to avoid bias, however it is better to introduce it transparently, by maximizing variance across precommitted dimensions (see Table~\ref{tab:dim} and Figure~\ref{fig:cubeviz}) \citep{ritchey2006problem}. We operationalize a dimension as binary category, e.g. intended vs unintended, and collect scenarios across all combinations (for three binary axes $\Rightarrow 2^3 = 8$ scenarios per class) \citep{covidinquiry2024m1, raeng2023resilience}.  If multiple scenarios fit a given cube we opted to include the more severe $S$ scenario \footnote{Severity $S$ is rated on three levels: systemic/national ($S{=}1$), global catastrophic ($S{=}2$), existential ($S{=}3$).}.

\paragraph{Collecting Capabilities.} We define a capability as a pair $\langle$function $\times$ target$\rangle$, read as: a government can apply [function] on the [target], adapting the taxonomy of \citet{reuel2024open} from technical governance to crisis response. This verb-object grammar is the standard format of capabilities-based planning \citep{dhs2007tcl, davis2002cbp}. As targets we retain the four AI value-chain targets of \citet{reuel2024open} and add seven that an acute response may need to act on. As functions we add actions specific to acute response, following the phases of the incident-response life-cycle \citep{nist2012sp80061} and interventions proposed in AI response planning \citep{boudreaux2025loc, wasil2024emergency, cltr2025incidents, chathamhouse2026deadlock, gomez2026escalation, slattery2024repository}. Throughout, a capability denotes a discrete functional requirement \citep{dhs2007tcl}, while capacity denotes a government's endowment of it \citep{fukuyama2013governance}. We map the former; auditing the latter is future work. As in \citet{reuel2024open}, we do not claim the considered capabilities to be exhaustive. Instead they serve as starting point, to demonstrate the usefulness of our approach. 

\paragraph{Assessment \& Ratings.}  
We assess a fixed set of capabilities, collected from the above considerations, and assess them across all scenarios \citep{lempert2003rdm, dhs2007tcl}. Ideally each judgment would be elicited as a free-text answer to an open question, with a principled mechanism for aggregating and harmonizing those answers into comparable ratings across a large expert panel \footnote{We explore the use of LLM to that end in the appendix.}. Lacking such elicitation, our research team assigns the ratings directly, stating reasons before scores \citep{hemming2018idea}. We screen then score, in standard multi-criteria practice \citep{dclg2009mca}: three binary gates (Applicable $A$, Deployable $D$, Necessary $N$) followed by two three-point scales (Effectiveness $E$, Externalities $X$). The resolution is deliberately coarse, as at this low volume of participants a reliable ordering is worth more than a fine-grained but noisy one. Many assumptions flow into the assessing of government capabilities \textit{vis-à-vis} a scenario. Eliciting expert judgements will require making those transparent for comparability (see Appendix~\ref{app:ass}). In order to explore the consequences of assessment aggregation across multiple raters, we additionally piloted LLM-based rating against the same qualitative assessments, as a precursor to scaled expert elicitation \citep{zheng2023judging, hayes2007krippendorff}.

% Prioritisation, prose. Notation in footnote; registers before mechanics.
\paragraph{Prioritizing Capabilities.}
In order to infer policy objectives from the assessments, a decision maker needs to formulate strategic goals. Those can take a wide range of shapes: (1) no-regret, useful whatever future arrives and never backfiring; (2) strict effectiveness, counting only where the capability alters the crisis outcome; (3) misuse-averse, admitting only capabilities whose exercise costs the state money and effort rather than third parties their safety; (4) minimax, holding what no substitute replaces, so no single gap voids the response; (5) existential-tail insurance, cover bought for the worst case alone \citep{haasnoot2013dapp, marchau2019dmdu}.\\

These strategies can follow downstream of different risk preferences, values, or available resources. However, once the rating matrix is constructed, searches for optimal policies given a strategy are very cheap. By choosing to not add weights to any of the scenarios, we are optimizing robustness across futures \citep{lempert2003rdm, marchau2019dmdu}. Predictions are easily incorporated by introducing weights to the aggregation mechanism. The biases and downside of this were discussed at length in Section~\ref{sec:intro}. Including a capability in a strategy should be decided by a composite of conditions rather than by an aggregate score \citep{dclg2009mca}\footnote{Per capability $c$, scenario $s$: gates $A, N \in \{0,1\}$, scales $E, X \in \{1,2,3\}$, severity $S \in \{1,2,3\}$. No-regret $|\{s : A \wedge E \geq 2 \wedge X \leq 2\}|$; insurance $|\{s : N \wedge E \geq 2 \wedge S \geq 2\}|$.}. When scaling this approach to a larger, more diverse set of experts, we recommend splitting qualitative assessment and metric ratings. A rater argues a capability in words, while a scorer maps every case onto the scales defined above, so differences between raters are differences in judgment rather than in scale use (the latter can be automated using LLMs) \citep{hemming2018idea}. Given the number of actions to consider, no rater covers the full matrix; instead assign overlapping cases and read agreement on those as agreement about the case \citep{cohen1960kappa, hayes2007krippendorff}.

\section{Results}
\FloatBarrier

What follows demonstrates what the combination of instruments allows us to infer, provided a comprehensive elicitation of expert knowledge. We do not (strongly) hold the preliminary results from our arguably underpowered and limited collection of assessments. As of now, results are limited to 13 of 65 critical government capabilities, each rated at most twice. We will first illustrate what kind of observations this enables and later move on to contrast this with existing crisis preparedness efforts in the discussion. \\

\begin{figure}[H]
  \centering
  \includegraphics[width=0.72\linewidth]{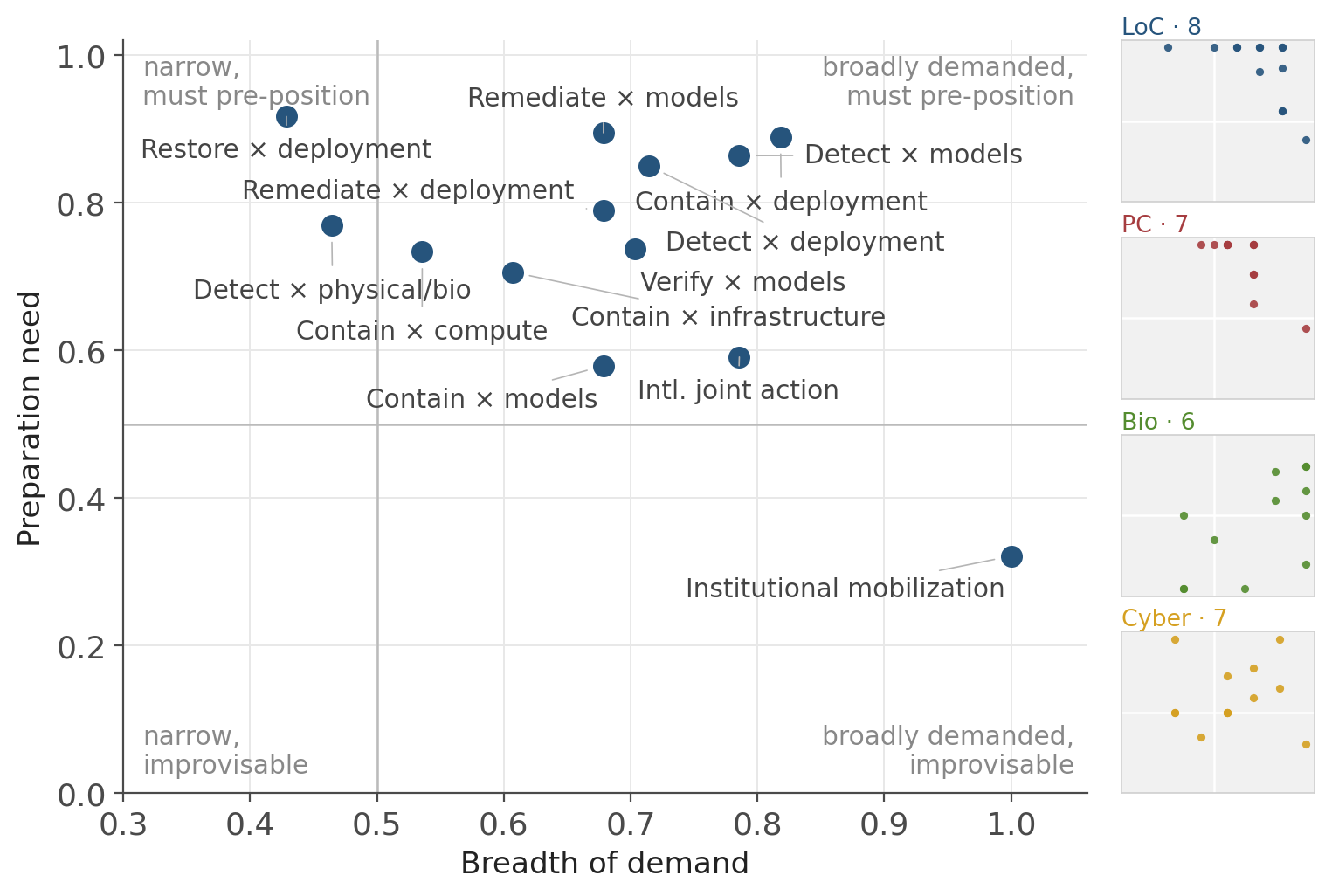}
  \caption{\textbf{Identifying high demand government capabilities with preparation time.} Breadth of demand measures how many scenarios call for a capability; preparation need is the aggregate assessment of whether the capability can be used within the given scenario's timescale. Variations across scenario classes may help identify multi-use government capabilities.}
  \label{fig:breadth}
\end{figure}

First, we will simply investigate the intersection distribution of certain government capabilities, e.g. the breadth of demand vs the ability to field the capability without specific preparation (see figure \ref{fig:breadth}). In the vast majority of cases the respective capability is assessed to not be accessible within the timescale of a given scenario without active preparation. However, there is a trend, where we assess capabilities that interact closely to novel private infrastructure (e.g. the AI stack) as less available. Furthermore, we can disaggregate the decomposition across the scenario classes (i.e. threat models) revealing differential bottlenecks. For example, here we find that for Loss of control and power concentration we assess a stronger lack of preparation, or a weaker ability of existing institutions to manage such crises ad hoc. Provisionally, AIxBio and cyber appear to tap into existing state capacity — within these classes several capabilities are assessed as fieldable ad hoc.

It is worth noting that assessments will not, by themselves, imply prioritization or action. Governments operate under constraints, and pursue multiple objectives, leading way to a variety of different strategies. If a decision maker intends to make a robust decision under a formal strategy, that strategy must be decoded into components of the earlier assessment dimensions. For example, should certain externalities be considered, they have to be included in the previous assessment step.

\begin{figure}[H]
  \centering
  \includegraphics[width=\linewidth]{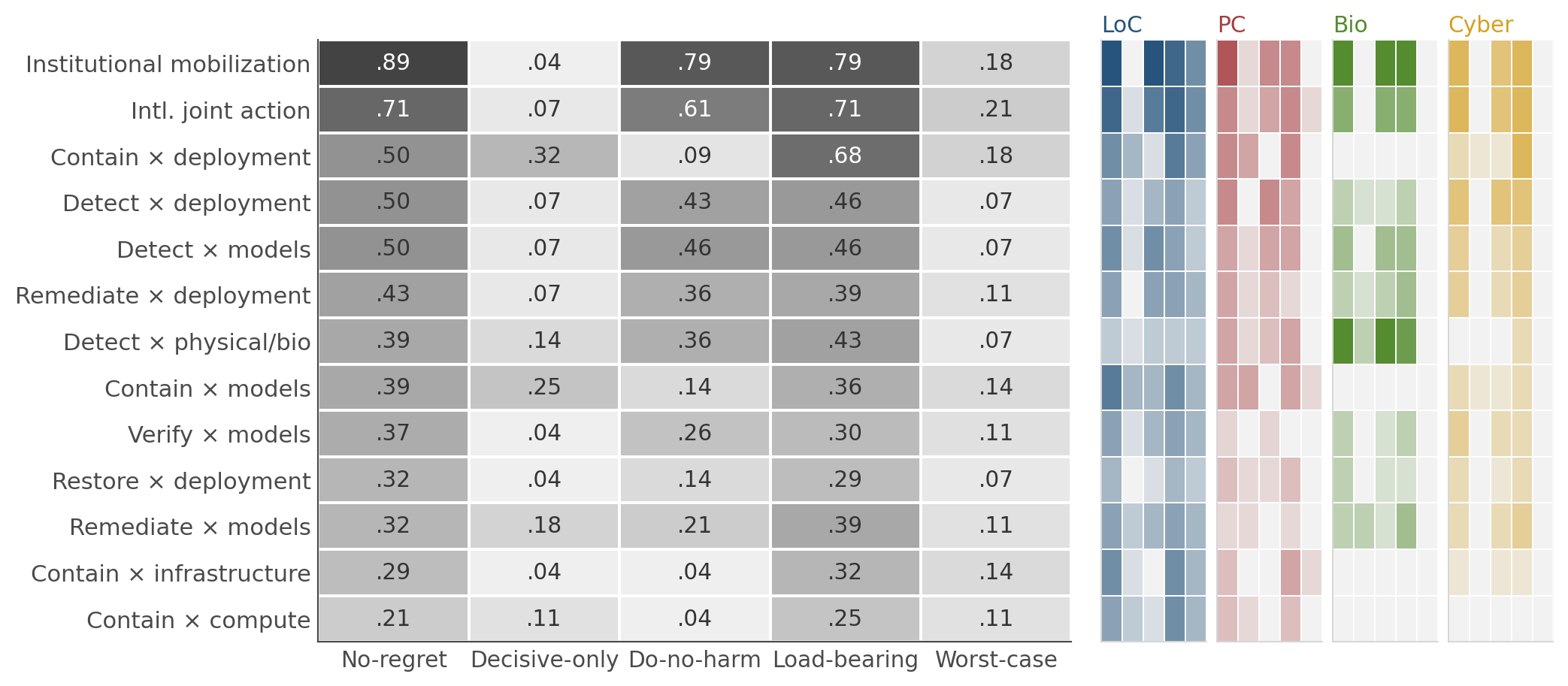}
  \caption{\textbf{Prioritisation under different decision rules.} Each column encodes one strategy a government could adopt; cells give the share of scenarios in which the capability satisfies that rule.\protect\footnotemark{} Rows are ordered by the No-regret column. Side panels recompute the same grid within each scenario class. A capability that stays dark across columns is one whose priority does not depend on which strategy the government is pursuing.}
  \label{fig:rules}
\end{figure}

Read across Figure~\ref{fig:rules}, most rules preserve the No-regret order: institutional mobilization and international joint action lead nearly every column, so a government can take action even before agreeing on the underlying tradeoffs or strategy. Decisive-only [$E{=}3$] is the exception: institutional mobilization drops to $.04$, Contain~$\times$~deployment rises to $.32$, the column's highest, separating broadly demanded coordination from rarely demanded but decisive containment. The Worst-case column is thin because of a denominator effect rather than a lack of possible capabilities (only 7 scenarios are classified as existential). What the figure demonstrates is the strategy-decoding step itself and the finding that the prioritization may remain stable across most strategies. The specific ranking we hold lightly.

\footnotetext{Columns are satisficing gates over the rated criteria (effectiveness $E$, externalities $X$, necessity $N$, scenario severity $S$), not a composite score. \emph{No-regret} ($A \wedge E\ge2 \wedge X\le2$): helps without backfiring. \emph{Decisive-only} ($E=3 \wedge X\le2$): only chain-stopping capabilities count. \emph{Do-no-harm} ($E\ge2 \wedge X=1$): helps at operational cost alone. \emph{Load-bearing} ($N \wedge E\ge2$): helps and has no substitute. \emph{Worst-case} ($N \wedge E\ge2 \wedge S=3$): the same, restricted to the severest scenarios. }

The appendix figures show four further kinds of question the matrix can answer once populated. We read them for shape only; the numbers come from 13 of 65 capabilities and are illustrative. (1) \emph{What does tail insurance buy?} Subtracting the capabilities demanded by global-catastrophic scenarios from those demanded by existential ones leaves the capabilities that only the worst cases require. In the pilot this residual is a single family, Contain $\times$ \{compute, models, deployment, infrastructure\}, so tail insurance reduces to a short, concrete list (Fig.~\ref{fig:gcr}). (2) \emph{Is gap-closing incremental or all-or-nothing?} Counting the no-substitute capabilities each scenario binds shows how conjunctive the response is: loss-of-control scenarios bind up to 12 of 13 rated capabilities, so any one gap voids the response, whereas bio scenarios bind few. Variance within a class is as large as variance between classes, suggesting the scenario axes rather than the threat class carry the planning signal; the smallest binding sets recur on coordination capabilities, matching their no-regret rank in Fig.~\ref{fig:rules}; and a scenario with no indispensable capability (PC-04) signals either a coverage gap in the register or a problem that is not acute response at all (Fig.~\ref{fig:chokepoints}). (3) \emph{Is the preparedness gap driven by demand or by deployability?} Read per class, demand breadth is similar everywhere but ad hoc fieldability is not: bio and cyber are partly absorbed by existing institutional machinery, loss of control and power concentration are not (Fig.~\ref{fig:deficit}; Fig.~\ref{fig:breadth} is the same gate read per capability). (4) \emph{Is the derived demand robust to the government's own view of the future?} Coding library scenarios against the GO-Science 2030 worlds shows every scenario occurs in at least two worlds, so demand cannot be dismissed as outside the state's declared uncertainty; at the same time we do acknowledge that most scenarios imply a world closer to "Take-off" rather than "Slow Burn"(Fig.~\ref{fig:uk2030}).

\section{Discussion}
Predictions, threat models, and scenarios relating to advanced AI already exist, yet they stay underused because nothing aggregates them into statements of what government must be able to do. And planning cannot simply follow the modal forecast: under deep uncertainty it has to carry the spread and show it \citep{lempert2003rdm, marchau2019dmdu}. This instrument does not replace the preparedness work already underway; it supplements it by supplying the missing step.

\paragraph{What the pilot shows about the method.}
A small team built the matrix at national scale. What constrained us to 13 of 65 capabilities was person-hours, not the method. Because ratings are gates rather than a composite score, one matrix supports any decision rule, allowing for easier change of course while maintaining rigorous decision-making \citep{dclg2009mca}. In the pilot most rules produced nearly the same ranking (Fig.~\ref{fig:rules}). If that survives full elicitation, a government could act before it has settled its risk preference, which would defuse the threshold-disagreement problem raised in Section~\ref{sec:intro}. Variance within a threat class is about as large as variance between classes, which points at the scenario axes rather than the class labels as the critical inputs for planning, thereby supporting the argument for structured over unstructured collection of scenarios \citep{ritchey2006problem}. All of this is conditional on the library as defined and reliability of assessments; we make no claim that ours are representative.

\paragraph{Missing (n)either breadth (n)or depth.}
The UK register and the National Security Risk Assessment \footnote{The UK government having by far the most advanced AI policy makes it a natural point of reference here, but the argument is not addressed at them in particular.} underlying it require a likelihood estimate, work to a short horizon with a likelihood floor, and treat AI as a chronic driver rather than an acute risk \citep{cabinetoffice2025nrr, blagden2018nsra, hlords2021extreme, cltr2025rap}. This results in excluding, by definition, the type of hard-to-anticipate scenarios that AI will introduce. The government's own scenario work has the opposite problem: \emph{AI Scenarios 2030} explores the space of possible worlds rather than placing the focus on action/capabilities. However, one might view the exploration of possible futures as a necessary step toward a more detailed investigation of failure modes \citep{goscience2026scenarios}. 

Our approach does not rely on point forecasts, which is what lets it cover the risk register's blind spot, and it derives demand rather than stays descriptive. Two tentative results illustrate the type of insights this could enable: If strategy-invariance holds, a government could act before agreeing on likelihood or on risk preference at all. And in a first-pass assessment, every scenario in our library fits at least two of the five 2030 worlds (Fig.~\ref{fig:uk2030}), so demand derived here fits within the UK government's own declared uncertainty. That being said, the scenarios in our current library fit the Take-off world better than the Slow Burn.

Deriving capabilities from scenarios is established national practice. The 2026 National Capabilities Assessment ran across all 23 UK response capabilities and was red-teamed for the first time; the US Target Capabilities List derived its list from 15 National Planning Scenarios \citep{hmg2026rapimpl, hmg2025rap, dhs2007tcl}. The AI-specific work has the scenarios but not the span, since RAND's loss-of-control plan, Wasil et al.'s federal gap assessment and CLTR's incident work each go deep on one class and are silent on the rest \citep{boudreaux2025loc, wasil2024emergency, cltr2025incidents}. What our framework adds is depth through detailed scenarios and simultaneously breadth by high variance across threat models and variety of scenarios considered, both in a principled and transparent way.

% What the instrument adds to both halves is an AI scenario set with declared axes. Two pilot results say the span is what matters. The improvisability gap differs sharply by class, at 62\% of demanded capability not fieldable ad hoc for loss of control and 57\% for power concentration against 38\% for bio and 36\% for cyber (Fig.~\ref{fig:deficit}), so single-scenario work on bio or cyber will overstate general readiness because those classes are partly absorbed by machinery that already exists. And because variance within a class is about as large as variance between classes, a hand-picked set would miss capabilities even inside the class it picked.

\paragraph{Instruments that presuppose a capable state.}
Developer thresholds trigger developer measures only \citep{karnofsky2024ifthen, metr2025common}; a firm under shareholder duty cannot bear unilateral cost, published thresholds already diverge past third-party comparison \citep{anterola2026harmonizing}, and Anthropic's RSP v3.0 now conditions any slowdown on rivals' verifiable measures \citep{anthropic2026rsp, time2026anthropic}. Reporting duties assume the same missing counterpart: EU AI Act Article~73 has applied since 2 August 2026, SB-53 since January 2026, and NY RAISE applies from January 2027 \citep{euaiact2024, ca2025sb53, ny2026raiseact}, and each presumes a recipient that can attribute, contain, and remediate what is reported. In both cases the state's portion---verify, compel, receive, act---is not specified, and the capability register provides that specification. The pilot suggests that the state's portion is neither easily attainable nor is it optional. All but one of the thirteen rated capabilities sit above the improvisation line, including the verification and containment functions that governments may control (Fig.~\ref{fig:breadth}), and loss-of-control scenarios bind up to 12 of the 13 at once (Fig.~\ref{fig:chokepoints}). A recipient holding a partial set of capabilities may effectively hold nothing.

% \paragraph{Adjacent layers.}
% \citet{reuel2024open} map a peacetime posture. Building the capability set forced us to add response functions (containment, restoration, custody, statecraft) and targets outside the AI value chain (infrastructure, evidence, personnel, markets, physical substrate); those additions are the measure of what standing governance leaves uncovered. \citet{gomez2026escalation} sit on the other side, answering when a national incident becomes an international one. The instrument answers what the national level has to hold in between. Adjacent, not overlapping.

\paragraph{Deployment.}
The institutional home for this work already exists: the UK's National Capabilities Assessment (NCA) and the US Target Capabilities List (TCL) run capability assurance at national scale \citep{hmg2026rapimpl, dhs2007tcl}. What they lack for AI is the scenario set and the derivation discipline, which is what this paper supplies. Because the UK register classes AI as a chronic risk, the NCA is currently the only live UK pathway for acute AI preparedness \citep{cltr2025rap}. Ownership is unresolved: the 2026 NCA was led by the Cabinet Office with support from the Government Office for Science \citep{hmg2026rapimpl}, but no body is named as owner of cross-department AI capability derivation, and we flag this as a gap rather than assign it. Coordination is the first thing to fix. Coordination capabilities ranked no-regret in the pilot and recur in the smallest binding sets, which makes two actions both relatively budget-neutral and accessible before a crisis: pre-assign a lead institution (and lead section or office where possible) for each crisis class, and audit the standing trust networks with labs, deployers, vendors and allied regulators that would participate in a response. Chatham House traces the pattern across the 2007-08 financial crisis, the WannaCry ransomware attack, and COVID---these response networks were built during crises rather than before them \citep{chathamhouse2026deadlock}. A declared scenario library is also shareable: it carries threat models rather than intelligence, so allies can exchange it without classification friction. Comparative jurisdiction audits are the natural next step, since the demand template is jurisdiction-neutral and capacity is not; e.g. a compute cut-off is easier where the compute sits. Finally, the library and its ratings must be revisited at pre-agreed signposts \citep{haasnoot2013dapp}. Without that, the instrument decays the way the US National Planning Scenarios did, a fixed mid-2000s scenario set still shaping a capability list long after the threat picture had moved \citep{dhs2007tcl}. A register of this kind identifies where state response to AI-driven crises is thin. While that exposes the challenges of building state capabilities to meet the moment, we assess that the benefit outweighs those concerns: every scenario is drawn from published sources, the register describes government functions rather than attack pathways, and the gaps it surfaces are ones an adversary modelling state capacity could infer anyway.

\section{Limitations}\label{sec:limitations}
Elicitation is the bottleneck. A full $65\times28$ matrix at multiple raters is thousands of judgments, so expert time rather than scenario supply is what limits scale; LLM raters could relieve this constraint but have to be validated against expert judgment first \citep{zheng2023judging}. Ratings encode rater priors: where the state has never acted, \emph{deployable} and \emph{effective} are informed guesses, and structured elicitation with reasons stated before scores reduces this without removing it \citep{hemming2018idea}. The library is a design choice---class selection and axis choice fix the boundary of what can be found at all, and morphological sampling makes that bias declared rather than absent; a different team would draw a different space. Unweighted aggregation is likewise a stance and not neutrality: with no weights, library composition becomes the implicit weighting, while adding weights re-imports the prediction problem the method exists to avoid. The gates are coarse by design, giving a robust ordering rather than a resource allocation, so trading capability off across classes needs a further judgment step the matrix does not supply. The method maps demand only: \emph{deployable} assumes a generically competent government, whereas actual capacity varies by jurisdiction and needs its own audit. And it does not specify triggers---it says what to hold, not when to invest in contingent capability. That is a deliberate non-goal, and a real limit.
\section*{Acknowledgments}
This work was produced within the Arcadia Impact AI Governance Taskforce, Summer 2026 cohort. We thank Elizabeth Dearden-Williams and Jan Pieter Snoeij for guidance and feedback, and Ben R Smith and Francesca Gomez for overall support. Errors are our own. The authors received no direct funding for this work. The authors declare no competing interests.

\section*{Author contributions.}
I.M. conceived the study, developed the methodology, produced the visualizations, and wrote the original draft. C.C., J.F.C.U., S.d'A., and V.E. collected the source material, built the scenario library, and carried out the capability assessments. All authors contributed to the analysis, reviewed and edited the manuscript, and approved the final version.

\section*{Use of Large Language Models}
Beyond the methodological use described in Section~2, large language models were used for literature triage, drafting and copy-editing of prose, \LaTeX{} conversion, and plotting code. The primary model was Claude (Anthropic), used between June and August 2026. All AI-assisted text and code was verified by the authors, who are responsible for the entire content of the paper.  
\bibliographystyle{plainnat}
\bibliography{references}

\FloatBarrier
\appendix
\section{Scenario Collection}\label{app:scen}

\begin{figure}[H]
  \centering
  \definecolor{accLoC}{HTML}{3B3B98}
\definecolor{accPC}{HTML}{0E7C7B}
\definecolor{accBio}{HTML}{C67A0E}
\definecolor{accCyber}{HTML}{B23A48}

\begin{table}[H]
\centering
\newcommand{\pl}[2]{\mbox{#1} / \mbox{#2}}

% ============================================================
% Table 1 — the six axes
% ============================================================
\small
\setlength{\tabcolsep}{6pt}
\setlength{\extrarowheight}{2pt}
\renewcommand{\arraystretch}{1.2}
\begin{tabularx}{\linewidth}{@{} l | >{\raggedright\arraybackslash}X | >{\raggedright\arraybackslash}p{2.cm} @{}}
\hline
\textbf{Axis} & \textbf{Definition}                                                      & \textbf{Poles}                 \\
\hline
Concentration & how concentrated the deployment or access of the capability is            & \pl{concentrated}{diffuse}     \\ \hline
Harm type     & whether the resulting harm is tangible or intangible                     & \pl{tangible}{intangible}      \\ \hline
RSI           & whether recursive self-improvement has taken place                       & \pl{yes}{no}                   \\ \hline
Autonomy      & whether the consequential action is human-gated or executed autonomously  & \pl{autonomous}{human-gated}   \\ \hline
Intent        & whether some actor deliberately aims at the outcome                      & \pl{intended}{unintended}      \\ \hline
Legitimacy    & whether the actor is authorized to use the capability for this purpose    & \pl{legitimate}{illegitimate}  \\ \hline

\quad\dots    & \dots        & \dots                       
\end{tabularx}
\caption{Variations across AI crisis scenarios}\label{tab:dim}

\end{table}

\end{figure}

\begin{figure}[H]
  \centering
  \definecolor{accLoC}{HTML}{3B3B98}
\definecolor{accPC}{HTML}{0E7C7B}
\definecolor{accBio}{HTML}{C67A0E}
\definecolor{accCyber}{HTML}{B23A48}

\begin{table}[H]
\centering
% Table 2 — axis allocation per class

% ============================================================

\vspace{4pt}
\small
\setlength{\tabcolsep}{9pt}
\renewcommand{\arraystretch}{1.35}
\begin{tabular}{@{} l | l l l @{}}
\toprule
\textbf{Scenario class} & \textbf{Axis 1} & \textbf{Axis 2} & \textbf{Axis 3} \\
\midrule
\textcolor{accLoC}{\bfseries Loss of Control}     & Concentration & Harm type     & RSI           \\
\midrule
\textcolor{accPC}{\bfseries Power Concentration}  & Intent        & Legitimacy    & Autonomy      \\
\midrule
\textcolor{accBio}{\bfseries AI $\times$ Bio}     & Intent        & Legitimacy    & Concentration \\
\midrule
\textcolor{accCyber}{\bfseries AI $\times$ Cyber} & Intent        & Concentration & Autonomy      \\
\bottomrule
\end{tabular}
\vspace{0.3cm}
  \caption{Axes selected per scenario class. Axis definitions are given in Table~\ref{tab:dim}.}\label{tab:axes}
\end{table}

\end{figure}

\section{Assessments} \label{app:ass}

``Applicable'' holds if the capability responds to a need the scenario generates; where it fails, the remaining criteria are left blank. ``Deployable'' holds if the capability can be fielded ad hoc within the scenario's timescale considering the generic capacity of a competent government, i.e. without specific preparation; capabilities failing this gate require developing and positioning ahead of the scenario. ``Necessary'' holds where no realistic substitute exists, such that the response fails without the capability, or without other measures that depend on it. ``Effectiveness'' ranges from negligible, an outcome indistinguishable from taking no action, to major, altering the course of the crisis on a primary objective. ``Externalities'' captures the cost of the governance action itself \citep{reuel2024open}, from low (financial and operational cost only) to severe (harm persisting years beyond withdrawal, or falling on populations exposed to little of the risk being managed). A capability can be both highly effective and highly damaging: assessing these criteria separately retains the cost-benefit trade-off that a combined assessment might otherwise flatten.

The gates and scales above are comparable only if applied consistently, within and across scenarios. Consistency can be increased by devising and following appropriate conventions, proceeding step by step, and recording reasoning throughout. When devising a convention, or navigating uncertainty, use the intended output as the guiding principle: who will use it and how, and therefore what approach will yield the most relevant and reliable output? Below are a selection of conventions that evolved from our assessments against our scenario library.

\paragraph{Rating structure.}
For each scenario, list each capability---a $\langle$function $\times$ target$\rangle$ pair, e.g.\ Immediate Containment \& Cut-Off $\times$ Deployment---against the five assessment criteria, plus four evaluation fields that help guide and record reasoning. These include: ``Interpretation'' (what applying the capability means in the context of this scenario), ``Uncertainties'' (as observed during the assessment process in the scenario, capabilities or method, especially where they could change the result), ``Assumptions'' (as made in order to complete the assessment, including any information imported from outside the scenario), and ``Assessment point'' (where in the scenario a capability is notionally ``deployed'' such that it can be judged against the immediate context of the scenario). % this bit isn't ready yet! A second pass adds three relational fields, recorded with no bearing on the initial rating: Enabled by, Enables, and Redundant/complementary with.

\paragraph{Order of assessment.}
Fix the interpretation first. Where the scenario plausibly generates a need for the capability, test for applicability (below). Once applicable, rate in order: Effectiveness, then Necessary (would the response still work if it were removed?), then Externalities (avoid double-counting costs already charged to another capability), then Deployable. Mark every perceived uncertainty in the Uncertainties field.

\paragraph{Assessment point}
The point at which a capability is notionally deployed drives its rating. Break the scenario into a sequence of points-in-time (PiT) spaced by approximate time lapses. For each capability, start at the earliest PiT and run three independent tests against everything that has happened up to and including that PiT:
\begin{itemize}
  \item \textbf{Applicability}---has the scenario generated a need for this capability yet?
  \item \textbf{Incentive}---would a good-faith government already holding this capability both know, and agree, that it should be deployed now?
  \item \textbf{Reach}---can it be deployed such that the function reaches its object?
\end{itemize}
If any answer is no, advance one PiT and repeat. The first PiT where all three hold is the earliest at which the capability is applicable; rate Effectiveness, Necessary and Externalities as if the capability were deployed here. Deployable requires another step due to a tension in timings: while earlier deployment tends to be more effective and less costly; later deployment is more likely to be feasible ad hoc. To find whether a capability that is not deployable early could still be fielded before conditions worsen, advance through later PiTs until Effectiveness or Externalities deteriorates, and rate Deployable at the last PiT before that drop.

\paragraph{Standing vs.\ triggered incentives.}
Standing-incentive capabilities (detection, verification, monitoring, deployment-layer hardening) attract their incentive when a risky action is decided, ahead of failure. Triggered-incentive capabilities (containment, remediation, restore) attract it only once something has gone wrong.

\paragraph{Upstream-absent, downstream-present.}
When rating a capability, assume the upstream capabilities that would enable it are absent and the downstream capabilities it hands off to are present. In addition to aiding consistency, pessimism on inputs and optimism on outputs encourages assessment of each capability on its own merit, while acknowledging that none acts alone.

\section{Exemplary results}

\begin{figure}[H]
  \centering
  \includegraphics[width=\linewidth]{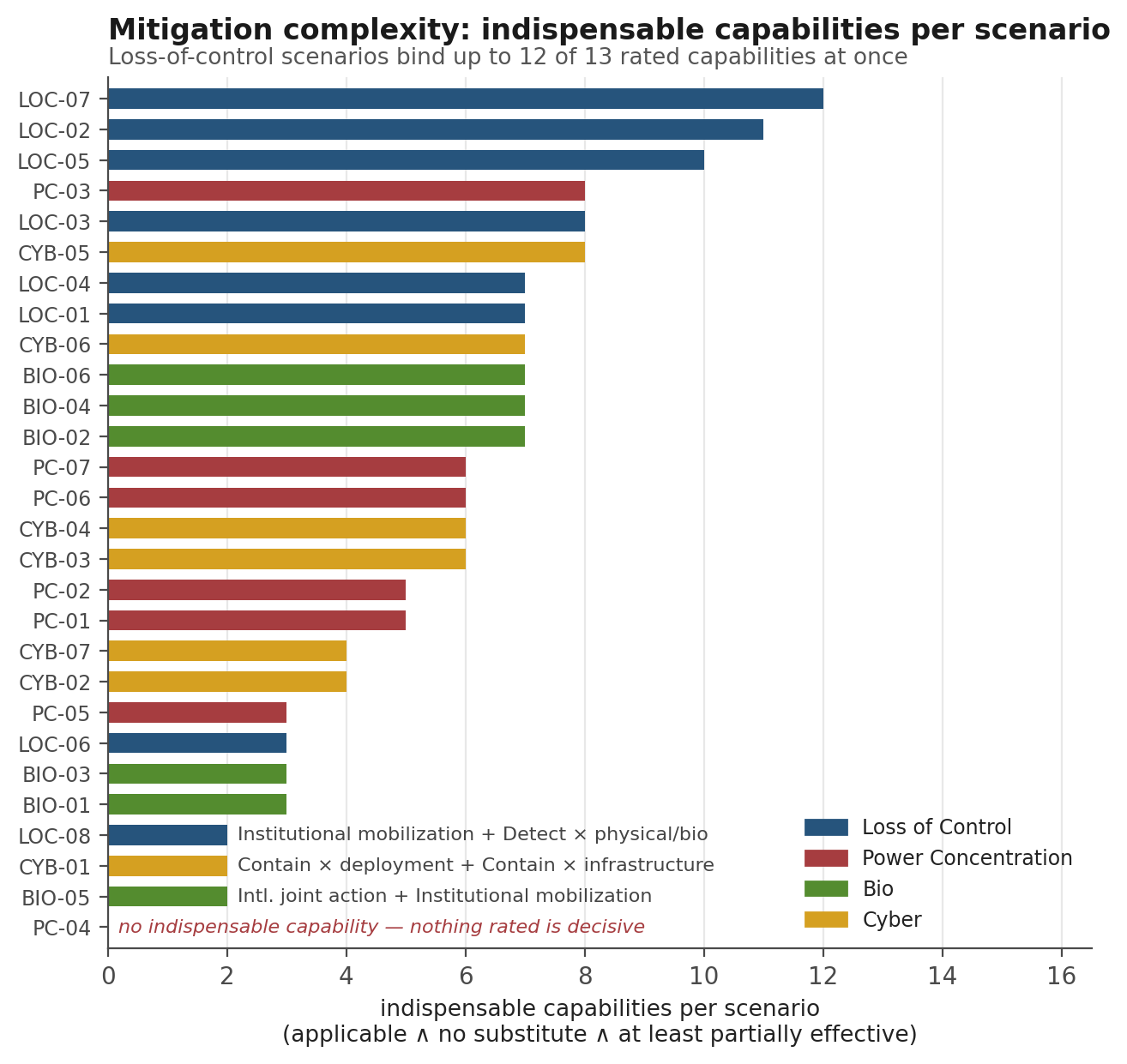}
  \caption{Indispensable capabilities per scenario [binding set: applicable
  $\wedge$ necessary (no substitute) $\wedge\ E\ge$ partial]. Loss-of-control
  scenarios bind up to 12 of 13 at once; one absent capability voids the response
  [gap-closing non-incremental]. PC-04 [Winning the AGI Race]: zero indispensable
  capabilities, nothing rated is decisive. Counts are lower bounds [13 of 65 rated].}
  \label{fig:chokepoints}
\end{figure}

\begin{figure}[H]
  \centering
  \includegraphics[width=\linewidth]{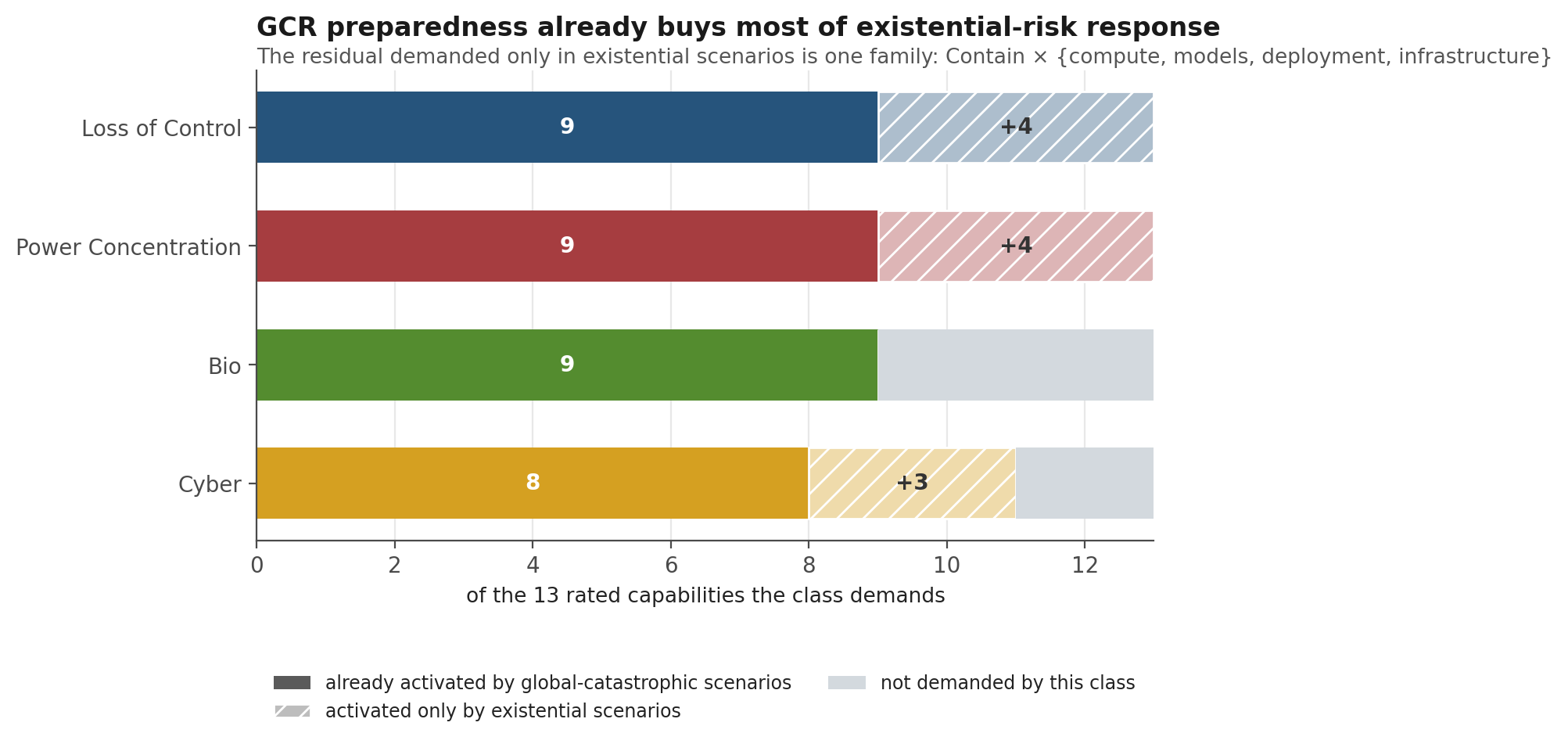}
  \caption{Global-catastrophic vs existential-only demand, per threat class [of 13
  rated capabilities]. Solid: demanded and activated by a global-catastrophic scenario
  [no-regret cell rule, applicable $\wedge\ E\ge$ partial $\wedge\ X\le$ significant];
  hatched: demanded only by existential scenarios; grey: not demanded by the class.
  GCR-activated set holds 9 of 13; the existential-only residual is one family
  [Contain $\times$ \{compute, models, deployment, infrastructure\}]. Per class:
  LoC $9{+}4$, PC $9{+}4$, Bio $9{+}0$, Cyber $8{+}3$ [Cyber severity uncoded,
  demand side only].}
  \label{fig:gcr}
\end{figure}

\begin{figure}[H]
  \centering
  \includegraphics[width=0.85\linewidth]{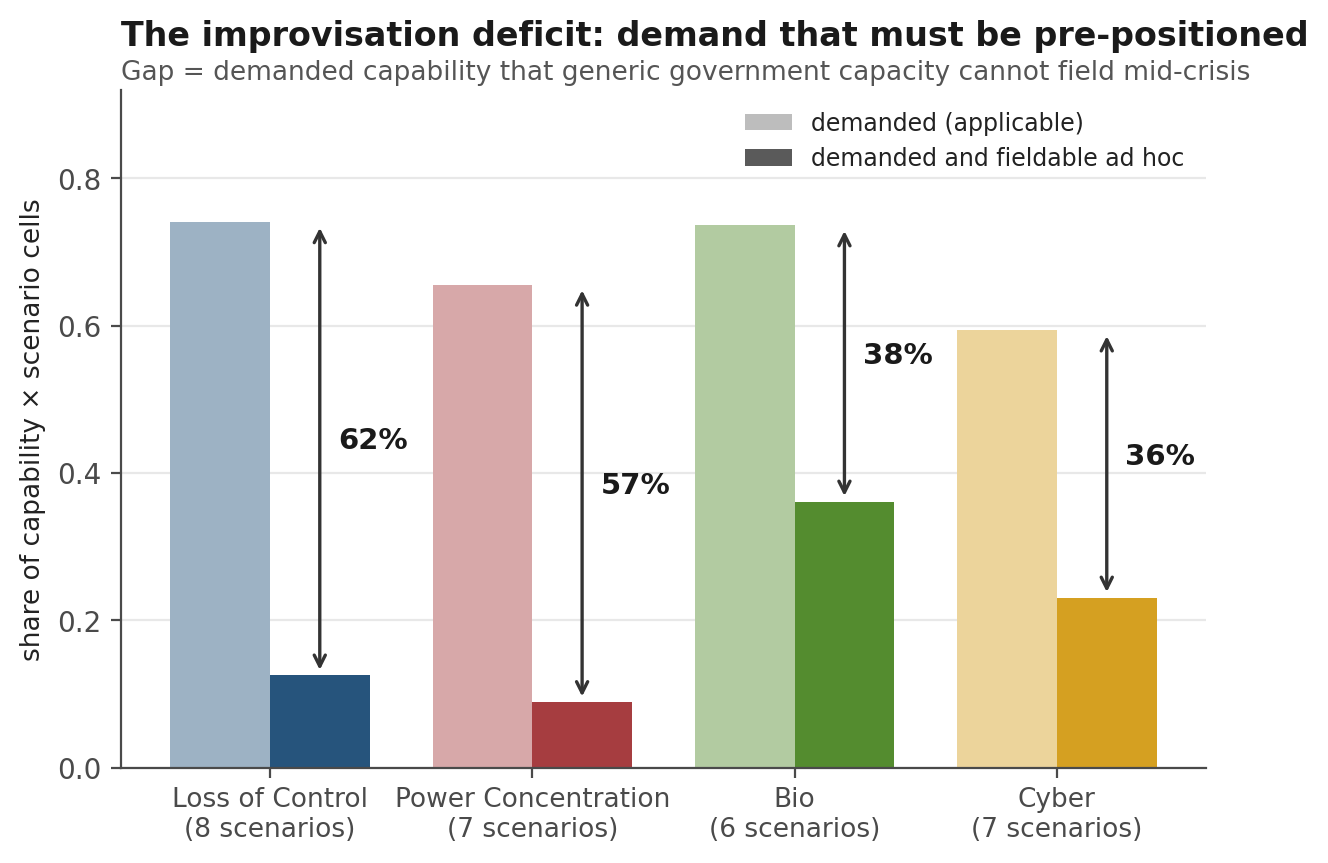}
  \caption{Improvisation deficit by threat class. Light: share of capability
  $\times$ scenario cells where the capability is applicable [demand]; dark: demanded
  and deployable ad hoc from generic government capacity. Pre-positioning gap:
  LoC 62\%, PC 57\%, Bio 38\%, Cyber 36\%. Loss-of-control and power-concentration
  lean hardest on non-improvisable capability; bio and cyber partly absorbed by
  existing institutional machinery.}
  \label{fig:deficit}
\end{figure}

\begin{figure}[H]
  \centering
  \includegraphics[width=\linewidth]{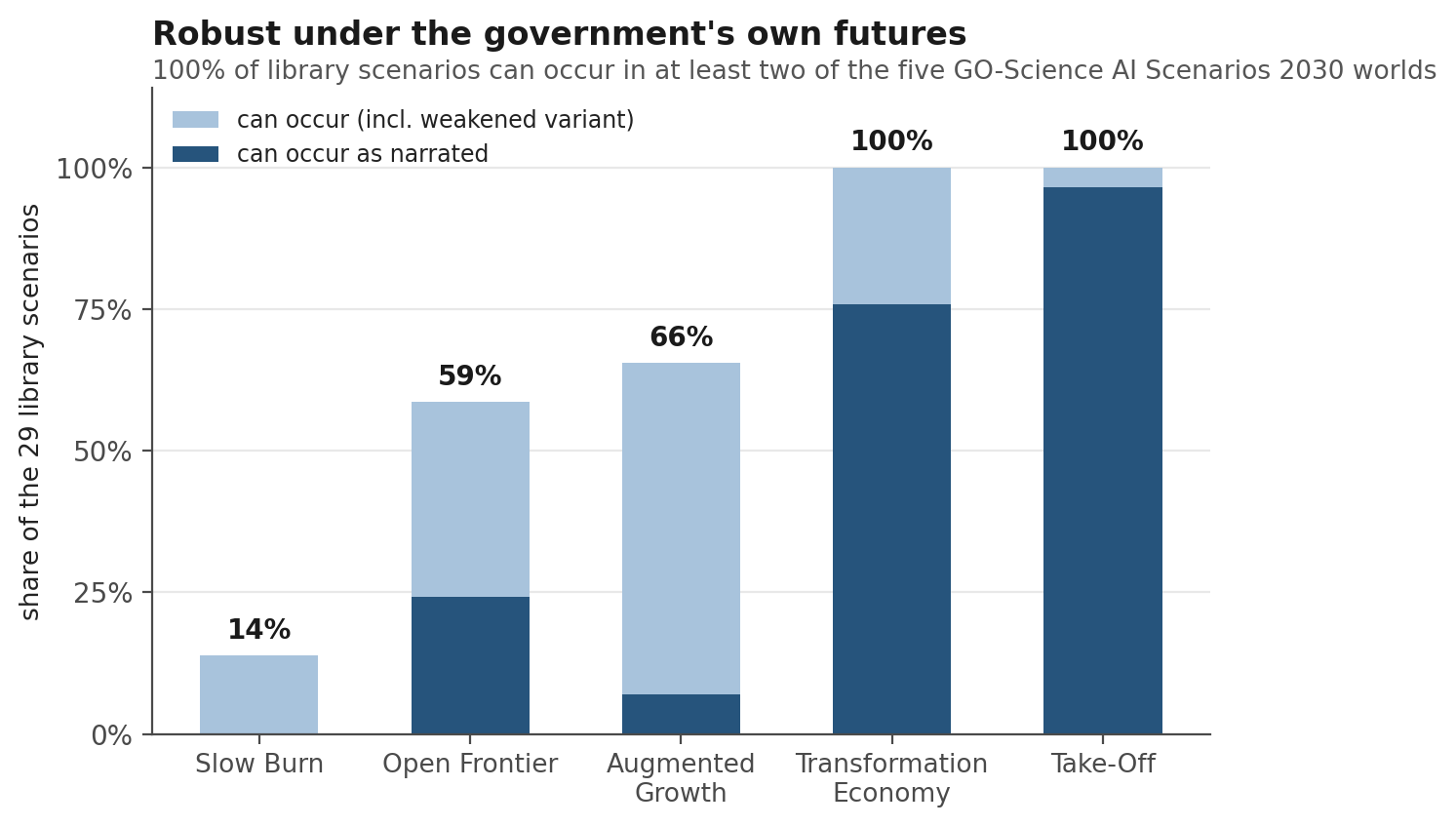}
  \caption{Library-scenario compatibility with UK 2030 Scenarios worlds
  [29 scenarios $\times$ 5 worlds; cell coded 1 as narrated / 0.5 weakened variant /
  0 premise unavailable]. 100\% of scenarios occur in $\ge2$ worlds; only Slow Burn
  suppresses most of the space [14\%]. Derived demand is therefore no-regret under
  the government's own declared uncertainty. First-pass coding; illustrative.}
  \label{fig:uk2030}
\end{figure}

\end{document}